\documentclass[twocolumn,showpacs,amsmath,amssymb,prl,superscriptaddress,floatfix,aps]{revtex4}

\usepackage{graphicx}% Include figure files
\usepackage{dcolumn}% Align table columns on decimal point
\usepackage{bm}% bold math

\usepackage{amssymb}
\usepackage{amsmath}
\usepackage{amsfonts}
\usepackage{amssymb}
\usepackage{color}

\usepackage{xcolor}
\definecolor{red}{rgb}{0.7,0,0}%darkred MIT
\definecolor{green}{rgb}{0.,0.35,0.}%darkgreen
\definecolor{blue}{rgb}{0.2,0.2,0.7} %beamer@blendedblue
\definecolor{black}{rgb}{0.15,0.15,.15}%not too black

\usepackage{hyperref}

\begin{document}

%\makeindex

\title{Trimer liquids and crystals of polar molecules in coupled wires}

\date{\today}
\author{ M. Dalmonte}
\affiliation{Dipartimento di Fisica dell'Universit\`a di Bologna and INFN, via Irnerio 46, 40127 Bologna, Italy}
\affiliation{IQOQI and Institute
for Theoretical Physics, University of Innsbruck, A-6020 Innsbruck, Austria}
\author{P. Zoller}
\affiliation{IQOQI and Institute
for Theoretical Physics, University of Innsbruck, A-6020 Innsbruck, Austria}
\author{G. Pupillo}
\affiliation{IQOQI and Institute
for Theoretical Physics, University of Innsbruck, A-6020 Innsbruck, Austria}
\affiliation{ISIS, IPCMS and UniversitŽ de Strasbourg, Strasbourg, France}

\begin{abstract}
We investigate pairing and crystalline instabilities of bosonic and fermionic polar molecules confined to a ladder geometry. Combining analytical and numerical techniques, we show that gases of composite molecular dimers as well as trimers can be stabilized as a function of the density difference between the wires. A shallow optical lattice can pin both liquids, realizing crystals of composite bosons and fermions. We show that these exotic quantum phases are robust against conditions of confinement of the molecular gas to harmonic finite-size potentials.
\end{abstract}


\maketitle

Binding among three or more particles plays a fundamental role in various physical systems ranging from quark confinement in QCD to tetramer bi-exciton formation in carbon nanotubes~\cite{matsunaga2011}; in quantum magnets it is responsible for the formation of many-body quantum phases of spin-multipoles~\cite{sudan2009}. In atomic quantum gases, two-body pairing is at the heart of paradigmatic phenomena, such as the observation of the BEC-BCS crossover~\cite{bec_bcs} in mixtures of ultracold fermionic atoms and the prediction of FFLO~\cite{FFLO} and Sarma~\cite{sarma} phases in spin-unbalanced gases. The study of few-body pairing, however, is usually confined to Efimov-like resonances~\cite{ferlaino10}, while  the realization of finite-density liquids made of composite particles is hindered by losses due to three-body recombination. Polar molecules confined to low-dimensional geometry provide a new opportunity to study inter-molecular pairing mechanisms and the associated quantum phases in a setup where collisional losses and also chemical reactions are suppressed~\cite{shape,miranda2011,PolMolExp}. Pairing of two spin-polarized fermionic molecules across coupled two-dimensional (2D) layers~\cite{wang2006} or one-dimensional (1D) wires~\cite{Kollath} has already lead to the prediction of, e.g., 2D inter-layer superfluidity~\cite{shlyapnikov2010} for the special case where the number of molecules is the same in all layers (wires). However, the pairing dynamics in the general situation where this number can vary across the layers (wires), as it happens in experiments~\cite{miranda2011}, has so far remained largely unexplored. Different particle numbers in the layers lower the overall (spin-rotational) symmetry of the problem, which makes the underlying physics drastically different~\cite{burovski}: a general pairing mechanism may exist for the formation of stable multi-molecule composite structures. The goal is now to determine whether (spin-rotational) symmetry breaking induces stable multimer liquids in confined quantum gases, where few-body processes are usually associated with resonances and losses~\cite{ferlaino10}.

\begin{figure}[t]{
\begin{center}
\includegraphics[width=0.95\columnwidth]{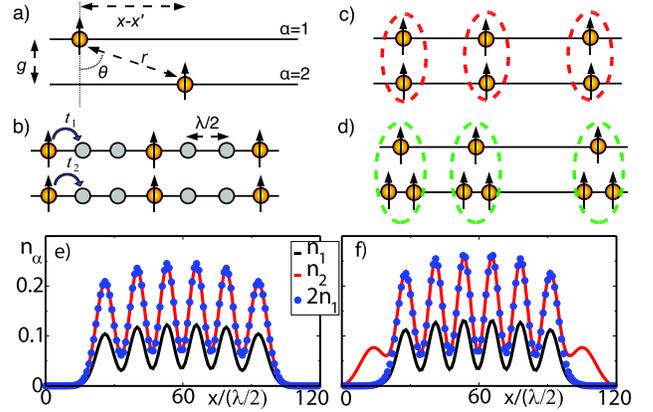}
\caption{(color online). (a): Molecules with dipole strength $d_\alpha$ ($\alpha=1,2$) in a two-leg ladder with interwire distance $g$ and dipolar interactions $V(r)= d_1 d_2(1-3 \cos^2\theta)/r^3$. (b-d): Sketch of phases (see text):  (b) dimer crystal in a lattice with spacing $\lambda/2$ and hopping energy $t_\alpha$; (c) dimer liquid; (d) trimer liquid. (e-f): Numerical results for density distributions $n_{\alpha}$ vs position $x$ in the presence of an optical lattice along the wires [see panel (b)] and weak harmonic confinement, showing trimers. Parameters [see also Eq.~\eqref{eq:Ham} and text]: $g=\lambda/2$, hopping asymmetry $t_2/t_1=0.01$, $d_1/d_0=3$, $d_2/d_0=0.5$ with $d_0^2=t_1 g^3$; the curvature of the harmonic potential is $\mathcal{K}/t_1=0.03$ (see text). Particle numbers $N_\alpha$: (e) $N_2=2N_1=12$; (f) $N_2=2N_1+2=14$; as a reference, blue points denote $2n_1$. The trap center is at $x/(\lambda/2)=60$.}
 \label{fig:fig1}
 \end{center}
 }
\end{figure}

In this letter we study a situation where polar molecules are confined to two coupled 1D wires, under conditions where the number of molecules can be the same or vary across the wires, similar to current experiments. This presents several new scenarios: (i) when the population is identical in the wires, a two-body bound-state is always present and is responsible for the appearance of dimer liquids, similar to 2D~\cite{shlyapnikov2010}, see Fig.~\ref{fig:fig1}(c). However, we find that (ii) {\it few-body} pairing is favored for {\it any} ratio of populations between the wires $P=p/q$, with $\;p,q\in\mathbb{N}$, which is a dense set between $]0,1]$. This leads to the stabilization of novel quantum phases of interacting fermionic or bosonic composite particles, where repulsive intra-wire interactions ensure collisional stability. In particular, we prove that a gas of trimers made of two particles on one wire and one in the other may be stabilized in these systems, Fig.~\ref{fig:fig1}(d). Both dimer and trimer liquids may be pinned by a weak period potential commensurate with the density, leading to a Luttinger-staircase of quasi-1D composite crystals~\cite{DPZ}. By numerical simulations, we show that these exotic quantum phases are robust to conditions of trapping in parabolic finite-size potentials and number fluctuations in the wires: unusual "wedding-cake" structures are formed where heavy composite particles are flanked by lighter one, Fig.~\ref{fig:fig1}(e-f).

Our starting point is the Hamiltonian $H=\sum_\alpha H_\alpha + H_{12}$ for the dynamics of molecules in the configuration of Fig.~\ref{fig:fig1}, with $\alpha=1,2$ and $H_\alpha$ the single-wire term
\begin{eqnarray}
H_{\alpha} & =& \int dx\;  \psi_{\alpha}^{\dagger}(x)\left[-\frac{\hbar^2 }{2m_{\alpha}}\partial_x^2+U_{\alpha}(x)\right]\psi_{\alpha}(x) \\
&+&\frac{d_{\alpha}^2}{8 \pi} \int dx \:dx' \frac{1}{|x-x'|^{3}}n_{\alpha}(x)n_{\alpha}(x') \nonumber \label{eq:Ham},
\end{eqnarray}
and $H_{12}=(d_1d_2)/(8 \pi) \int dx \:dx' V(x-x') n_1(x)n_2(x')$ the inter-wire coupling, with $V(x-x')=[1-3\cos^2(\theta)]/[g^2+(x-x')^2]^{3/2}$ showing a short-distance inter-wire attraction.
Here, $m_{\alpha},d_{\alpha}$ are the mass and the dipole strength respectively, and $\psi_{\alpha}$ ($\psi_{\alpha}^{\dagger}$) are fermionic or bosonic annihilation (creation) operators, with $n_{\alpha}(x)=\psi^{\dagger}_{\alpha}(x)\psi_{\alpha}(x)$; $U_{\alpha}(x)=\mathcal{U}_{\alpha}\sin^2(2\pi x/\lambda)$ represents an underlying periodic potential, as usually provided by an optical lattice~\cite{bloch_review} with wavelength $\lambda$ and depth $\mathcal{U}_{\alpha}$; $\theta$ is the angle between particles in different wires, with distance $g$, Fig.~\ref{fig:fig1}. The 1D Hamiltonian $H$ is valid for interparticle distances $n_\alpha^{-1} \gg (d_\alpha^2/\hbar \omega_\perp)^{1/3}$ with $\omega_\perp$ the frequency of transverse confinement provided by, e.g., a 2D optical lattice. For LiCs, RbCs and KRb molecules with $d_1=5.6$, 1.25, and 0.5 Debye, respectively, and $\omega_\perp = 2 \pi \times 100$ kHz, $(d_\alpha^2/\hbar \omega_\perp)^{1/3}$ is of the order of 360, 130 and 70 nm~\cite{shape}. In addition, $d_1 d_2/g^3 < \mathcal{U}_{\alpha}$, \cite{Foot}. Similar setups may be obtained on chip-based microtraps~\cite{Santambrogio2011}.

In~\cite{DPZ} it is shown that in the absence of an optical lattice ($U_{\alpha}=0$) and of inter-wire interactions, the dynamics in each wire is described by an effective Tomonaga-Luttinger liquid (TLL) theory with Hamiltonian~\cite{bosonization}
\begin{equation}
\mathcal{H}_{\alpha}=\frac{\hbar v_{\alpha}}{2\pi}\int dx\left[(\partial_{x}\vartheta_{\alpha}(x))^{2}/K_{\alpha}+K_{\alpha}(\partial_{x}\phi_{\alpha}(x))^{2}\right].\nonumber
\label{eq:eqLL}\end{equation}
Here $v_{\alpha}$ and $K_{\alpha}=(1+0.73n_{\alpha}R_{\alpha})^{-1/2}$ are the effective sound velocity and the TLL parameter, respectively~\cite{DPZ}, with $R_{\alpha}=m_{\alpha}d_{\alpha}^2/(2\pi\hbar^2)$ the intra-wire dipole length, and $\vartheta_{\alpha},\phi_{\alpha}$ represent long-wavelength density and phase fluctuations~\cite{bosonization}. For finite interactions, Eq.~\eqref{eq:Ham} is then effectively described by $\mathcal{H} = \sum_{\alpha}\mathcal{H}_{\alpha}+\mathcal{H}_{12}$, with $\mathcal{H}_{12}=\mathcal{H}_{f}+\mathcal{H}_{b}$. Here, in the weak coupling regime we obtain $\mathcal{H}_{f}=-\frac{d_1d_2}{24g^2\pi^3} \int dx \partial_x\vartheta_1(x)\partial_{x}\vartheta_2(x)$
for the {\it quadratic} forward-scattering part of the interactions~\cite{bosonization}, by approximating the interwire interaction with its zero-component Fourier transform\cite{bosonization}; $\mathcal{H}_{b}$ is the {\it back scattering} part, with a typical sine-Gordon-type (sG) form, to be discussed below. The effect of $\mathcal{H}_{f}$ is to modify the effective TLL parameters, while $\mathcal{H}_{b}$ can induce novel pairing instabilities. Here, we first discuss the case of two identical ({\it balanced}) coupled wires, and then the more general case of ({\it unbalanced}) wires with different densities, molecular masses and interactions.

\emph{In the balanced case}, the quadratic part $\sum_\alpha \mathcal{H}_\alpha+\mathcal{H}_f$ of $\mathcal{H}$ can be diagonalized by introducing standard charge and spin fields, $\vartheta_{c,s}=(\vartheta_1\pm\vartheta_2)/\sqrt{2}$~\cite{epaps}, leading to a description in terms of coupled TLLs with effective parameters $K_{c,s}$. For weak interactions we estimate $K_{c,s}=K_1\left[1\mp\Gamma_{12}K_1/v_1\right]^{-1/2}$, $\; \Gamma_{12}=d_1^2/(24\pi^2\hbar g^2)$. The term $\mathcal{H}_{b}$ has the sG form
$\mathcal{H}_{b}\propto-\frac{n_1^2d_1d_2}{12g^2\pi}\int dx \cos[2\sqrt{2}\vartheta_s(x)]$,
and, in agreement with Berezinskii-Kosterlitz-Thouless (BKT) theory~\cite{bosonization,Haller2010}, it is relevant, thus opening a spin gap (e.g., pairing), if $K_s<1$; this condition is always satisfied and shows that, similar to the 2D case, pairing of molecules across the wires is always favored in 1D, even for an infinitesimally small attraction between the wires with $g,R_\alpha\ll n_\alpha^{-1}$.
\begin{figure}[t]{
\begin{center}
\includegraphics[width=0.9\columnwidth]{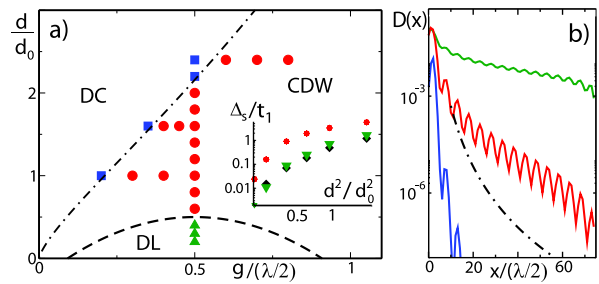}
\caption{(color online). (a): Phase diagram in the balanced case with $d_\alpha=d$, $t_2=t_1$ and $n_\alpha=0.2$ on a lattice with spacing $\lambda/2$ and $d_0^2=t_1 (\lambda/2)^3$. Green triangles, red circles and blue squares are numerical results for dimer liquid (DL), charge-density wave (CDW) and dimer crystal (DC) phases, respectively; dashed and dashed-dotted lines are qualitative phase boundaries. Inset: scaling of the dimer-pairing spin-gap $\Delta_s$ as a function of $d^2$ for several example densities $n_\alpha$ and interwire distances $g$: [$n_\alpha=0.2,g=0.35\lambda$] (red points), $[0.4,\lambda/2]$ (green triangles) and $[0.2,\lambda/2]$ (black diamonds). (b): Dimer correlation function $\mathcal{D}(x)$ vs $x$ in the phases of panel (a), with $g/(\lambda/2)=0.5$, $n_\alpha=0.2$. Top to bottom (continuous lines): $d/d_0=0.5$ (DL), 1.3 (CDW) and 2.2 (DC); the dashed-dotted line is the CDW-DC quantum phase transition at $\sim x^{-25/4}$.  }
 \label{dimers_pd}
 \end{center}
 }
\end{figure}
The role of dipolar interactions is evident in the charge sector: in contrast to models with attractive contact interactions such as the Hubbard model~\cite{bosonization}, here $K_c$ can be much smaller than 1; as an example, in the strongly interacting regime $n_{\alpha}R_{\alpha}\gg1$, where dimers are well approximated by tightly-bound composite particles with effective mass $M=2m$ and dipole strength $D=2d$, $K_c \simeq 2/(1+5.8 n_\alpha R_\alpha )^{1/2}$. As a result, the many-body groundstate shows a crossover from a dimer liquid (DL) with dominant pair correlations $\mathcal{D}(x)=\langle \psi^{\dagger}_{1,i}\psi^{\dagger}_{2,i}\psi_{1,i+x}\psi_{2,i+x}\rangle\sim |x|^{-1/K_c}$ for $K_c> 1$ to a charge-density wave (CDW) with dominant density correlations  $\mathcal{G}(x)=\langle n_in_{i+x} \rangle \sim |x|^{-K_c}$ for $K_c< 1, n_i=n_{i,1}+n_{i,2}$. In this regime, the CDW can be pinned by a very shallow optical lattice commensurate with the particle density, stabilizing a  Luttinger staircase\cite{DPZ} of dimer crystals (see \cite{epaps} for more details).

We verify numerically these predictions in the deep lattice regime $\mathcal{U}/E_r\gg 1$, with $E_r$ the lattice recoil, where an appropriate description is given in terms of an anisotropic extended Hubbard model (AEHM)~\cite{LewensteinReview}:
\begin{eqnarray}\label{ham_lattice}
\hat{H}&=&-\sum_{\alpha,i}t_{\alpha}(c^{\dagger}_{\alpha, i}c_{\alpha, i+1}+h.c.)- \frac{2d_1d_2}{g^3}\sum_i n_{1,i}n_{2,i}+\nonumber\\
%&+& \sum_{\sigma,i<j}\frac{d_{\sigma}'^2}{(j-i)^3}n_in_j+d_1'd_2'\sum_{i<j}V_{ij}(n_{1,i}n_{2,j}+n_{2,i}n_{1,j})\nonumber
&+&\sum_{i<j}\left[d_1d_2V_{ij}(n_{1,i}n_{2,j}+n_{2,i}n_{1,j})+\sum_{\alpha}\frac{d_{\alpha}^2n_{\alpha,i}n_{\alpha,j}}{(j-i)^3} \right],\nonumber
\end{eqnarray}
which we analyze using a quasi exact Density-Matrix-Renormalization-Group technique (DMRG)~\cite{schollwock2005}.
Here, $V_{ij}$ describes the anisotropic part of the dipolar interaction, $t_{\alpha}=1$ sets the energy scale. In the balanced case, $d_\alpha =d$ and $n_\alpha=n$. The field theoretical description of the AEHM in terms of continuum fields is equivalent to the one in the limit of a shallow lattice given above, and we thus expect a similar qualitative behavior.

Figure~\ref{dimers_pd} shows the phase diagram for a commensurate density $n_\alpha = 0.2$. By fixing $g=\lambda/4$ and increasing $d$, we find first a crossover from a TLL of dimers (DL) to a CDW, and then a BKT-type pinning quantum phase transition to a DC with $n_\alpha = 1/5$ (phase boundaries are discussed in \cite{epaps}). Examples of $\mathcal{D}(x)$ are plotted in panel (b) for all three phases, where the dash-dotted line marks the transition from power-law to exponential decay.
We also calculate the spin gap $\Delta_s$, by performing a finite-size scaling of $\Delta_s(L)=E_L(N, N)-E_L(N+1,N-1)$, with $E_{L}(M,M')$ the ground state energy at finite size $L$ in the sector with $n_1=M,n_2=M'$, for different densities and $g$ [Inset of panel (a)]. We find that a finite gap is present in the entire phase diagram, as expected, although it is small for weak interactions due to the BKT scaling $\Delta_s\propto \exp[-\beta d^2]$,~\cite{bosonization}.

\emph{The unbalanced case} presents unconventional instabilities. As shown in~\cite{burovski}, the Haldane expansion of density operators in a  two-component TLL generates a infinite series of massive terms coming from different combinations of vertex operators
\begin{equation}
\mathcal{H}_b= \sum_{p,q\in \mathbb{N}}G_{p,q}\int dx  \cos[2x(pk_{F1}-qk_{F2})+2(p\theta_1-q\theta_2)],\nonumber
\end{equation}
where $G_{p,q}$ are model dependent coefficients and $k_{F\alpha}=\pi n_{\alpha}$. With the exception of the simplest case $p=q=1$, these terms are usually negligible from a renormalization group point of view. However, we find that strong dipolar interactions drastically enhance the effect of $\mathcal{H}_b$~\cite{epaps}, allowing the formation of \emph{multiparticle composites}~\cite{burovski}, or \emph{multimers}, made of $p/q$ particles on the upper/lower wire respectively. In particular, composite objects with $p=1$ may become particularly stable, corresponding to a population ratio $n_2/n_1=\kappa\in \mathbb{N}$; in such a situation, analogous to the above discussion of dimers, a term in $\mathcal{H}_b$ with the sG form $G_{\kappa, 1}\int dx \cos[2(\kappa\vartheta_1-\vartheta_2)]$ may become relevant and stabilize a \emph{multimer liquid}, uniquely identified by the finite gap associated with the bound state formation and an algebraic decay of multimer correlations $\langle \mathcal{M}^{\dagger}(0)\mathcal{M}(x)\rangle, \quad \mathcal{M}=(\psi_1)^{\kappa}\psi_2,$
while the single-particle and dimer correlations $\mathcal{D}(x)$ decay exponentially. However, qualitative estimates  (see \cite{epaps}) indicate that comparatively large interaction strengths are needed in order to stabilize such a liquid, and the critical strength increases with increasing $\kappa$. Thus, below we focus on $\kappa = 1$ and we investigate numerically the possibility to realize a trimer liquid (TL) by considering the AEHM  with both interaction and hopping asymmetry, quantified by the ratios $d_2/d_1$ and $t_2/t_1$.
\begin{figure}[t]{
\begin{center}
\includegraphics[width=0.9\columnwidth]{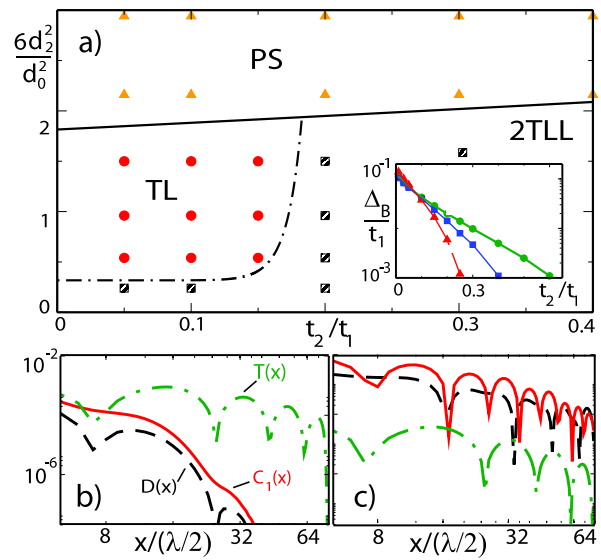}
\caption{(color online). (a): Phase diagram in the unbalanced case, as a function of $d_1^2/d_0^2$ and $t_2/t_1$. Here, $d_1=6d_2$, $g=\lambda/2$, $n_2=2n_1=0.1$ and $d_0^2=t_1 (\lambda/2)^3$; circles, triangles and squares are numerical results for the trimer liquid (TL), phase separation (PS) and two independent Tomonaga-Luttinger-Liquid phases (2TLL), respectively; lines are guides for the eye. Inset: Trimer binding energy $\Delta_B/t_1$ vs $t_2/t_1$ for several choices of $(d_1/d_0; d_2/d_0)$: triangles, squares and circles are (0.5; 3), (0.3; 1.8) and (0.5, 2), respectively. (b-c): Correlation functions for trimers $\mathcal{T}(x)$ (dashed-dotted line), independent TLL $\mathcal{C}_1(x)$ (continuous line) and dimer liquid $\mathcal{D}(x)$ (dashed line) vs $x$ in a system of size $L=120$ (see text). (b): Correlations in a TL with $d_1/d_0=3$, $d_2/d_0=0.5$, $t_2/t_1=0.1$; (c) 2TLL with $d_1/d_0=2.4$, $d_2/d_0=0.4$, $t_2/t_1=0.4$.}
 \label{fig:fig3}
 \end{center}
 }
\end{figure}

To get a guidance on possible trimer instabilities, we first compute the binding energy of the three body problem, $\Delta_B=\lim_{L\rightarrow \infty}[E_L(1,1)+E_L(0,1)-E_L(1,2)]$. Based on a Born-Oppenheimer approach~\cite{petrov07}, we expect a sizeable binding in the regime $t_1 \gg t_2$ and $d_1 \gtrsim d_2$, where a fast particle in wire $1$ binds two repulsive heavy molecules in wire 2. Numerical examples of $\Delta_B$ vs $t_2/t_1$ are given in the Inset of Fig.~\ref{fig:fig3}, for a few values of $d_2,d_1$ and $g=\lambda/2$, as trimers are unfavored for too large or too small interwire distances~\cite{epaps}. In particular, for strong attraction with $g\ll \lambda$, the formation of a dimer plus an unpaired particle is favored (see also the discussion of phase-separation (PS) below). We have further investigated the phase diagram in the low-density limit ($n_2=2 n_1=0.2$ and 0.1) as a function of $d_1,\; d_2$ for system sizes up to $L=120$ using DMRG. The TL phase is characterized by an exponential decay of both $\mathcal{D}(x)$  and single particle correlation function $\mathcal{C}_{\alpha}(x)=\langle c^{\dagger}_{\alpha,i}c_{\alpha,i+x}\rangle$, whereas the trimer correlator $\mathcal{T}(x)=\langle c^{\dagger}_{1,i}c^{\dagger}_{2,i}c^{\dagger}_{2,i+1}c_{2,i+x+1}c_{2,i+x}c_{1,i+x}\rangle$ decays algebraically. This is in contrast to the case of two coupled TLL, where no correlation is exponentially suppressed. The phase diagram for $d_1=6d_2,n_2=0.1$ is plotted in Fig.~\ref{fig:fig3} (see also~\cite{epaps} for the dependence on $d_1/d_2$): the TL extends in a broad region, and survives even for comparatively small interaction strength and interaction asymmetry, albeit in both cases small hopping rates $t_2\leq 0.2$ are needed. Sample correlation functions for both TL and 2TLL are plotted in Fig.\ref{fig:fig3}(b-c). For large interactions $d_1d_2/g^3\gg t_{\alpha}$, {\it microscopic} phase separation (PS) can occur. The latter corresponds to the formation of a gas of strongly-bound, mutually repulsive dimers across the wires, coexisting with a gas of unpaired molecules in wire 2, see~\cite{epaps}.

\textit{Effect of the trap}. We have investigated the fate of these exotic trimer bound states in the lattice under conditions of harmonic trapping in finite-size potentials with Hamiltonian $\hat{H}_{t}=(\mathcal{K}/L^2)\sum_i n_i(L/2-i)^2$, with $\mathcal{K}$ the curvature. We characterize the TL phase in inhomogeneous trapped systems via the local order parameter $\delta n_i=2n_{1,i}-n_{2,i}$, describing the deviations between the density of the majority and minority components~\cite{epaps}. Trimers exist if $\delta n_i\ll 1$. We find that trimers are remarkably robust to the presence of high-density, finite-size effects, and fluctuations in particle numbers in the wires. The general situation is one where composite particles with a larger mass (e.g., trimers) occupy the central region of the trap, and are flanked by lighter particles, being dimers or single excess particles. This determines an unusual "wedding-cake" structure in the density profile, reminiscent of that observed with cold atom with contact interactions~\cite{bloch_review}. Figures~\ref{fig:fig1}(e-f) show example results of the density profile in the trap, in the presence of composite trimers for a case of an exactly-matching number of particles in the two wires, and a situation where the particle number in wire 2 is larger by 20 percent. In both cases, we find a well defined region at the trap center where trimers are formed, with $\delta n_i \sim 10^{-8}$.

Significant asymmetry in dipoles and mass is obtained in mixtures of, e.g., LiCs and RbCs molecules trapped in independent optical lattices, where a weak optical lattice in the wire direction provides additional tuning of the (effective) mass. Alternatively, single-species molecules can be prepared in different internal (e.g., rotational, vibrational) states with different dipole moments~\cite{miranda2011,micheli07}. For a 1D lattice in the wire direction, the effective mass is tuned via internal-state-dependent tensor shifts~\cite{micheli07,friedrich97}. All phases discussed above can be detected via the measurement of decay of correlation functions [see, i.e., Fig.~\ref{fig:fig3}(b-c)], using, e.g., direct {\it in situ} imaging techniques~\cite{Weit2011}. Particle correlations across the wires will be detected via {\it in situ} imaging~\cite{Weit2011} as well as noise correlation measurements~\cite{bloch_review}. In addition, the TL, DL and DC may be spectroscopically probed~\cite{burovski}. For example, molecular dimers with sizeable, spectroscopically resolvable, gaps $\Delta_s \gtrsim 1\mu$K are obtained by trapping, e.g., LiCs molecules in a deep 2D optical lattice with spacing $\lambda/2=400$, trapping frequency $\omega_\perp = 2 \pi \times100$ kHz.

The realization of the above scenario in polar molecule experiments will lead to the first observation of trimers in the many-body context with cold gases.

We thank E. Burovski and A. Chotia for discussions, E. Ercolessi and F. Ortolani for discussions and for providing the DMRG code. This work was supported by MURI, AFOSR, EOARD, IQOQI, the Austrian FWF, the EU through NAME-QUAM, COHERENCE and AQUTE.

{\it Note:} after completion of this work, we became aware of the related work~\cite{wunsch} on few-body boundstates of polar molecules in coupled tubes.

\newpage

\section{Supplementary material}

\subsection{Luttinger staircase of dimer crystals}

In the balanced, strongly interacting regime, where CDW correlations dominates over DL ones, the system may undergo a BKT phase transition towards a crystalline, fully gapped phase in presence of an optical lattice. When the periodic potential lattice spacing $\lambda/2$ is a integer multiple of the mean interparticle distance $1/n$, that is  $n\lambda/2=1/f, f\in\mathbb{N}$, the optical lattice is properly described by a sine-Gordon term\cite{bosonization}:
\begin{equation}
\mathcal{H}_{OL}^{bc}\propto\frac{\mathcal{U}_{\alpha}n_1^2d_1d_2}{g^2}\int dx \cos[2f\sqrt{2}\vartheta_c(x)],
\end{equation}
where the coefficient $\mathcal{U}_{\alpha}$ is proportional to the depth of the lattice. According to BKT scaling, a crystalline phase is stabilized if the TLL parameter of the charge sector is smaller than a critical value $K_c^{(c)}=4/f^2$\cite{bosonization}, thus forming a Luttinger staircase in analogy with the single tube case~\cite{DPZ}. Each DC is characterized by a periodic structure with one dimer every $f$ sites and is fully gapped.

\subsection{Balanced case: numerical results}

\begin{figure}[h]{
\begin{center}
\includegraphics[width=7cm]{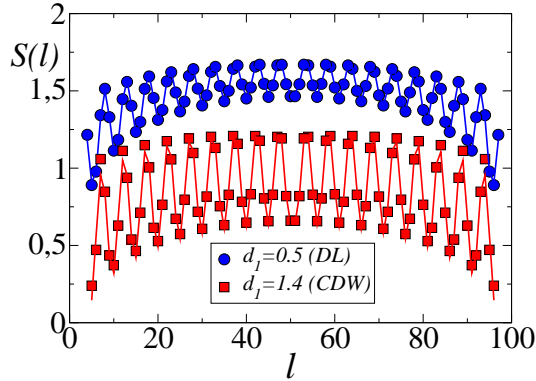}
\caption{(color online). Entanglement entropy of a subblock of length $l$ with respect to the system for $d=0.5, 1.4$: numerical datas (blue points and red squares respectively) and fits to CFT prediction \cite{calabrese2010sup} which give a central charge $c=0.981, 0.97$, in good agreement with the predicted value for a gapless phase phase with $\Delta_s\neq 0$.}
 \label{bal_BE}
 \end{center}
 }
\end{figure}

\begin{figure}[h]{
\begin{center}
\includegraphics[width=6.5cm]{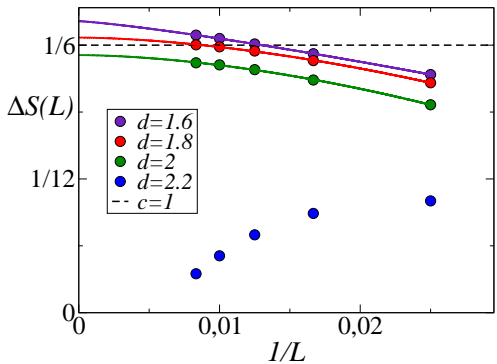}
\caption{(color online). Scaling of $\Delta S$ as a function of the system size; points denote numerical datas, whereas thick lines are fits. The dashed line correspond to $c=1$. For $d<2.2$, the asymptotic is in good agreement with a $c=1$ gapless phase, whereas it goes to zero in the DC phase.}
 \label{deltaS}
 \end{center}
 }
\end{figure}

In this section, we provide additional informations about the numerical calculation in the balanced case. The first transition of interest is the
CDW-DC one, which corresponds to the pinning of composite bosons by the underlying lattice. In both regions, $\Delta_s>0$, whereas the charge sector is gapless in the CDW and gapped in the DC. A clear signature of such a transition is given by an exponential decay of $\mathcal{D}(x)$; however, due to the large exponent of such a correlation close to the transition line, we choose to apply an alternative method based on the investigation of the so called \emph{block entropy} (BE):
\begin{equation}
S(l)=- Tr \rho_A \log_2 \rho_A
\end{equation}
where $\rho_A$ is the reduced density matrix of a block $A$ of length $l$. For critical model subject to open boundary conditions, and described by a conformal field theory (CFT), it was shown that (see \cite{calabrese2010sup} and references therein):
\begin{equation}
S_L(l)=\frac{c}{6}\log_2 (l) + {\rm const}+ \mathcal{O}(1/l)
\end{equation}
where $c$ is the so called central charge, $L$ is the system size and  $l$ the cord distance. In our case, DL and CDW phases are described by a CFT with $c=1$ (see Fig. \ref{bal_BE}), whereas the DC is a non-critical, fully gapped phase whose block entropy saturates to a constant value over a certain distance. As proposed in \cite{lauchli2008sup}, the BE can be used to characterize a transition from critical to non-critical phase by computing
\begin{equation}
\Delta S(L)=S_L(L/2)-S_{L/2}(L/4).
\end{equation}
In the thermodynamic limit, this quantity saturates to $c/6$ for critical theories and to $0$ for non-critical ones: red points and blue squares in Fig. 2 reflect this property, as can be inferred from the numerical data presented in Fig. \ref{deltaS} . We notice that oscillations typical of Tomonaga-Luttinger liquids\cite{calabrese2010sup} do not allow for an exact estimate of $c$, even though the difference between critical and non-critical models is still manifest.

\begin{figure}[t]
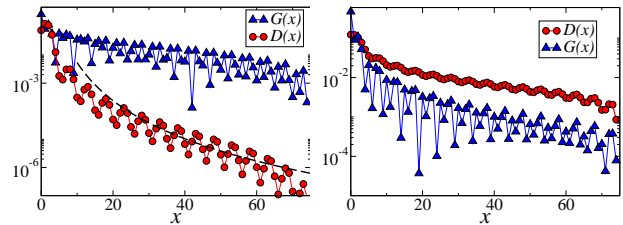
{
\begin{center}
\includegraphics[width=4cm]{sup_Fig3a.eps}
\includegraphics[width=4cm]{sup_Fig3b.eps}
\caption{(color online). Correlation functions $\mathcal{D}(x),\mathcal{G}(x)$ (red circles and blue triangles respectively) in the CDW (left panel) and DL phase (right panel) obtained from DMRG in a $L=100$ system. Here, $g=0.5$ and $d=0.5, 1.4$ respectively; the dashed line in the left panel denotes a reference curve $x^{-4.5}$.}
 \label{DL_CDW_corr}
 \end{center}
 }
\end{figure}

The DL-CDW transition is instead more subtle, as it corresponds to a precise value of the TLL parameter $K_c=1$. While an accurate determination of the phase boundary is involved, we just characterized point in the phase diagram by looking at $\mathcal{D}(x)$ and $\mathcal{G}(x)$. Sample numerical datas representing points close to the phase boundary are presented in Fig. \ref{DL_CDW_corr} .

\begin{figure}[t]{
\begin{center}
\includegraphics[width=6cm]{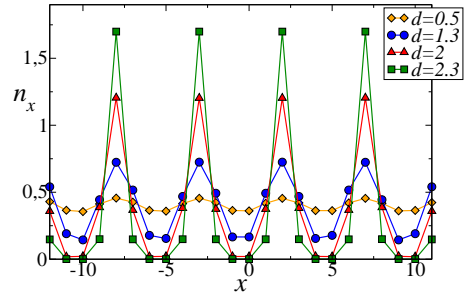}
\caption{(color online). Density distribution $(n_{1,i}+n_{2,i})$ in the balanced case for $n_1=n_2=1/5$; orange diamonds, blue circles, red triangles and green squares denote $d=0.5, 1.3, 2$  and $2.3$ respectively. The density distribution becomes sharper and sharper as we approach the DC phase.}
 \label{bal_dens_distr}
 \end{center}
 }
\end{figure}

Finally, it is worth mentioning that the density distribution changes drastically as we increase the interaction strength. In Fig. \ref{bal_dens_distr}, we notice that in the DC particles are locked to certain positions, whereas in both gapless phases the density shows modulations which become smaller in the DL phase.

%Contents: some detail about the balanced phase diagram, including entanglement entropy and CDW correlation functions.

\subsection{Unbalanced case} 
 
Unbalanced gases have been shown to give rise to a series of exotic pairing instabilities when the density ratios of the two component is a rational number\cite{burovskisup,roux2011}. In general, the TLL parameter of both components should be very small in order to establish such exotic bound states; in our particular case, one can estimate the TLL parameter of the $\kappa+1$ composite field $\vartheta_M=(\kappa\vartheta_1-\vartheta_2)/\sqrt{2}$ by neglecting effective spin-charge interactions, getting $K_M=(\frac{\kappa^2K_1+K_2}{2})(1+\kappa K_1K_2\chi/(\kappa^2K_1+K_2))^{-1/2}$, where we have considered equal velocities $v_1=v_2$ and a general forward scattering term  proportional to $g'=-v\chi$. This results recovers previous expressions presented in Refs.\cite{burovskisup,roux2011}, and can be extended to consider general density ratios $n_1/n_2\in\mathbb{Q}$. However, the stability of bound states with more that three particles is challenging; in fact, large density imbalances may favor phase separation or collapse over pairing. This intriguing perspective goes beyond the scope of the current work, in which only TLs were investigated via numerical simulations.
 
\subsubsection{Numerical Results}
 
\begin{figure}[t]{
\begin{center}
\includegraphics[width=6cm]{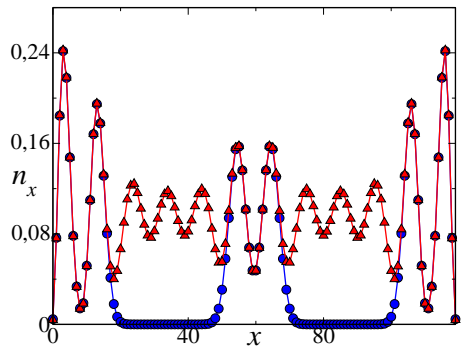}
\caption{(color online). Density distributions $n_1$(blue circles) and $n_2$(red triangles) in the phase separated state (see text).}
 \label{unbal_PS}
 \end{center}
 }
\end{figure}

We discuss now some additional numerical results on the specific unbalanced case $n_2=2n_1$. In Fig. \ref{unbal_PS} we plot the density distribution of both species in the strongly interacting regime, $d_1=5.6, d_2=0.9,t_2=0.1$, where we have fixed $t_1=1$ and $n_2=1/10$; DMRG calculations were performed in this regime by taking up to 512 states per block and 7 sweeps in $L=120$ chain. The system phase separates into small high density {\it islands} where $n_1\simeq n_2$ and regions where $n_1=0$. This phenomenon is justified by the fact that such strong interaction induces a strong repulsion between heavy particles, whereas the interwire interaction is strongly attractive, thus favoring two-particle pairing, as present in the islands.

\begin{figure}[t]{
\begin{center}
\includegraphics[width=6cm]{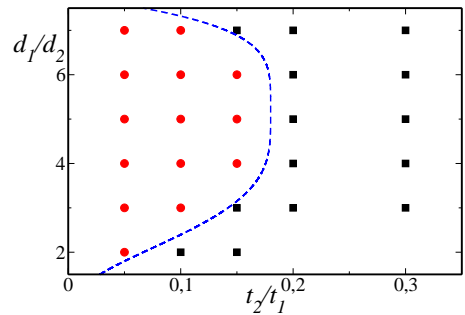}
\caption{(color online). Phase diagram in the unbalanced case $n_2=2n_1=1/10$ at fixed $d_2=0.5$: black squares and red circles denote 2TLL and TL phase respectively. The dashed line is a guide for the eye.}
 \label{unbal_pd}
 \end{center}
 }
\end{figure}

Let us turn our attention on the role of the interaction unbalance $d_2/d_1$. From the analytical treatment, it is difficult to extract quantitative informations about how large this quantity should be in order to realize a trimer phase. In Fig.\ref{unbal_pd}, we present the phase diagram of the system at a fixed interaction strength $d_2=0.5$ and different interaction asymmetry $1<d_2/d_1<8$. The TL phase survives even for relatively small interaction unbalance, albeit the critical hopping ratio $t_2/t_1$ becomes smaller and smaller as we approach $d_1=2d_2$.

\subsubsection{Density locking in the TL phase}

\begin{figure}[t]{
\begin{center}
\includegraphics[width=6.5cm]{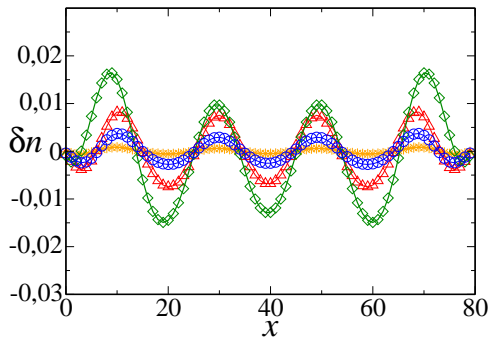}
\caption{(color online). Density locking $\delta n$ in the homogeneous case, calculated in a $L=80$ site chain. Green squares, red triangles, blue circles and orange stars represent different values of $t_2/t_1=0.3,0.2,0.1,0.05$, the other relevant parameters being $n_2=2n_1=1/20,d_1=3=6d_2$. }
 \label{hom_denslock}
 \end{center}
 }
\end{figure}

The choice of a \emph{ local effective order parameter} related to the TL is of great interest when the goal is to investigate an inhomogeneous situation such as a trapped gas. In case of multiparticle pairing, a natural choice, in analogy with the two-component case\cite{meisnersup}, is $\delta_{\kappa} n=\kappa n_1-n_2$, which in the trimer case becomes:
\begin{equation}
\delta n=2n_1-n_2.
\end{equation}
In Fig. \ref{hom_denslock}, we plot $\delta n$ for different values of $t_2/t_1$ with $d_1=3=6d_2$ and $n_2=2n_1=1/10$ in a $L=80$ sites chain; the oscillatory profile, due to open boundary conditions, display a maximum smaller than $\delta n_c\simeq 0.05$ when we are in the TL phase (blue and orange lines), whereas it's considerably larger for $t_2\geq 0.2$. 

\begin{figure}[t]
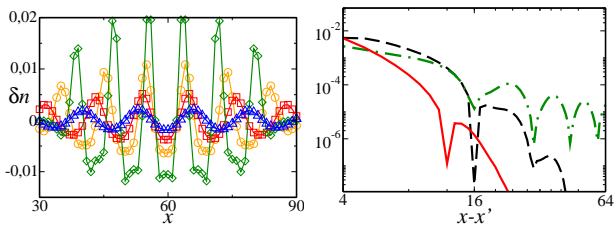
{
\begin{center}
\includegraphics[width=4cm]{sup_Fig8a.eps}
\includegraphics[width=4cm]{sup_Fig8b.eps}
\caption{(color online). Panel {\it a}: density locking in the inhomogeneous case: $\mathcal{K}=1$(green diamonds), 0.3 (black circles), 0.1 (red squares) and  0.03(blue triangles). Panel {\it b}: correlation functions in the inhomogeneous case, $\mathcal{K}=0.03$, in a $L=120$ site system, with respect to $x'=L/4$; $\mathcal{T}(x)$ (dot-dashed), $\mathcal{D}(x)$ (black) and $\mathcal{C}_1(x)$ (red). In both panels, $d_1=3=6d_2,t_2=0.1$.}
 \label{inhom_denslock}
 \end{center}
 }
\end{figure}

A direct comparison between homogeneous and inhomogeneous setup can be performed by focusing on the central part of the system. In Fig. \ref{inhom_denslock} (left panel), we plot $\delta n$ for a $L=120$ system around the trap minimum: for large values of the trapping potential, the density locking is not as tight as needed in order to stabilize a TL in the homogeneous case, whereas for more shallow traps, $\mathcal{K}\leq0.1$, the oscillations are compatible with a TL. As a further check, we considered correlation function decay in the central region of the system, plotted in Fig.\ref{inhom_denslock} (right panel) in the shallow trap regime; $\mathcal{C}$ and $\mathcal{D}$ correlations decay exponentially, only $\mathcal{T}$ being algebraically decaying, confirming the presence of a TL in the central part of the system.

\subsubsection{Coexistence of dimers and trimers}

In presence of slightly different density ratios with respect to exact commensurabilities the sine-Gordon term responsible for exotic pairing instabilities\cite{burovskisup, DPZ2sup} does oscillate, and thus bound states in general are not stable. However, the spatial-dependent potential drastically changes this picture. In fact, particles can rearrange in order to realize a commensurate phase in the middle of the system, expelling particles in excess towards the edges.

 \begin{figure}[h]{
\begin{center}
\includegraphics[width=7cm]{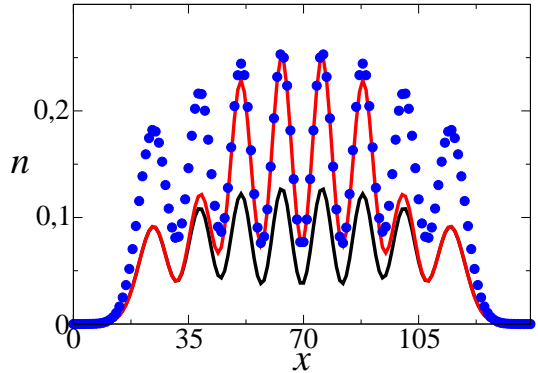}
\caption{(color online). Density distribution $n_1$(black line), $n_2$(red line) and $2*n_1$(blue points) for $\mathcal{K}=0.03,t_2=0.1, d_1=3=6d_2$ in a $L=160$ sites system, with  $N_1=9, N_2=14$; in the central part, $n_2\simeq 2n_1$, whereas away from the center $n_1\simeq n_2$.}
 \label{trap_coex}
 \end{center}
 }
\end{figure}

The $2n_1<n_2$ case is plotted in \cite{DPZ2sup}: in this case, trimers form in the middle, whereas extra heavy particles occupy both wings. In the opposite limit $2n_2>2n_1>n_2$ one can expect naively a similar picture, specially because of the much smaller mass of light particles with respect to trimers. However, when intermediate-to-large interactions are considered ($d_1*d_2\gtrsim t_1$), this configuration is energetically unfavored because of relevant interwire attraction which prevents light particles to occupy region of space where $n_2=0$. In Fig. \ref{trap_coex} we plot the density profile of both species for $\mathcal{K}=0.03,t_2=0.1, d_1=3=6d_2$ in a $L=160$ sites system with unbalanced, non commensurate densities, $N_1=9, N_2=14$. The trap center still presents a strong density locking of $2n_1-n_2$, whereas in the outer part the system resemble a DL where $n_1=n_2$; in this configuration, different types of paired phase coexist in the same experimental setup.

\end{document}